\documentclass[runningheads]{svjour2}
\smartqed  
\usepackage{graphicx, epsfig, amssymb} 
\usepackage{amsmath, amsfonts}
\usepackage{bm} 
\usepackage[linktocpage]{hyperref}
\usepackage[caption=false]{subfig}
\usepackage[usenames]{color}
\newcommand{\be}{\begin{equation}}
\newcommand{\ee}{\end{equation}}
\newcommand{\beq}{\begin{eqnarray}}
\newcommand{\eeq}{\end{eqnarray}}

\newcommand{\bea}{\begin{eqnarray}}
\newcommand{\eea}{\end{eqnarray}}
\newcommand{\ben}{\begin{enumerate}}
\newcommand{\een}{\end{enumerate}}
\newcommand{\bi}{\begin{itemize}}
\newcommand{\ei}{\end{itemize}}

\journalname{General Relativity and Gravitation}
\begin{document}

\title{Cosmic Censorship and parametrized spinning black-hole geometries}


\author{Vitor Cardoso        
\and
Leonel Queimada}


\institute{Vitor Cardoso \at
              CENTRA, Departamento de F\'{\i}sica, Instituto Superior T\'ecnico - IST,\\
              Universidade de Lisboa - UL, Av.~Rovisco Pais 1, 1049 Lisboa, Portugal \\
              \email{vitor.cardoso@ist.utl.pt}             \\
              \emph{Present address:} Perimeter Institute for Theoretical Physics, 31 Caroline Street North
Waterloo, Ontario N2L 2Y5, Canada
           \and
           Leonel Queimada \at
              CENTRA, Departamento de F\'{\i}sica, Instituto Superior T\'ecnico - IST,\\
              Universidade de Lisboa - UL, Av.~Rovisco Pais 1, 1049 Lisboa, Portugal \\
              \email{leonel.queimada@ist.utl.pt}            
}

\maketitle

\begin{abstract}
The ``cosmic censorship conjecture'' asserts that all singularities arising from gravitational collapse
are hidden within black holes. We investigate this conjecture in a setup of interest for tests of General Relativity:
black hole solutions which are parametrically small deviations away from the Kerr solution. These solutions have an upper bound on rotation, beyond which a naked singularity is visible to outside observers.
We study whether these (generic) spacetimes can be spun-up past extremality with point particles or accretion disks. Our results show that cosmic censorship is preserved for generic parameterizations.
We also present examples of special geometries which {\it can} be spun-up past extremality.
\end{abstract}


\maketitle

\section{Introduction}

The coming decade will see unprecedented advances in our knowledge of strong-field gravity phenomena.
Radio and deep infrared telescopes will probe the region close to the Schwarzschild radius of supermassive black holes
(BHs) -- in some cases producing direct images -- whereas gravitational-wave telescopes are expected
to finally make the first direct detection of waves generated during the inspiral and merger of two BHs or neutron-stars.
These observations will probe highly dynamic, relativistic events taking place in strong-field regions, thus offering a unique
test of General Relativity (GR) itself~\cite{Berti:2015itd}.
 
One of the most cherished beliefs is that astrophysical BHs are described by the Kerr geometry, characterized by
only two parameters, its mass $M$ and angular momentum $J=aM$~\cite{Teukolsky:2014vca}. This belief is borne out of uniqueness results in GR, and the fact that the Kerr geometry
is linearly and most likely nonlinearly stable~\cite{Dafermos:2010hd,Cardoso:2014uka} against massless fluctuations. Of interest for us here, is the additional property that Kerr BHs spin slowly, $a/M\leq 1$. 
For higher values of rotation a (naked) singularity becomes visible to asymptotic observers. It turns out that it is impossible -- at least in a perturbative regime -- to spin-up
the spacetime beyond the extremal $a=M$ value with point-like particles~\footnote{Point-like particles are a very useful proxy for other more complex and realistic scenarios, such as accretion.}~\cite{1974AnPhy..82..548W,BouhmadiLopez:2010vc,Barausse:2010ka,Barausse:2011vx,Colleoni:2015afa,Rocha:2011wp}, in accordance with the spirit of the Cosmic Censorship Conjecture (CCC)~\cite{Wald:1997wa}.
Even in fully nonlinear and extremely violent processes, such as the collision of two equal-mass Kerr BHs, the CCC seems to be preserved~\cite{Sperhake:2009jz,Sperhake:2012me,Cardoso:2014uka}.

However, natural extensions of GR as simple as a minimally coupled scalar or the addition of a Gauss-Bonnet invariant coupled to scalars
allow for other, more generic BH solutions~\cite{Berti:2015itd,Herdeiro:2014goa,Yunes:2011we,Pani:2011gy}. Thus, an important starting point to test GR is to understand the nature of compact objects and to devise tests that allow us to confirm or exclude the ``Kerr paradigm''~\cite{Ghasemi-Nodehi:2015raa,Jiang:2015dla,Bambi:2015rda,Moore:2015bxa}.

Tests of GR are usually performed within working hypothesis, for instance that some modified theory characterized by a
coupling $\gamma$ to other fields is a viable and interesting candidate. Then, observations are used to impose constraints on the coupling constant $\gamma$. Other approaches accept our ignorance on the ultimate theory of gravity
and simply parametrize the compact object by some unknown functions or constants, and use observations to constraint
the geometry of these parameterized compact objects~\cite{Berti:2015itd,Bambi:2015rda,Moore:2015bxa,Johannsen:2011dh,2013ApJ...773...57J,Bambi:2013eb,Cardoso:2014rha,Bambi:2015kza}.

Our goal is to understand how BH spacetimes behave. We restrict ourselves to a very simple class of problems: the infall of point particles into rotating BHs. On the other hand, we study this simple process in a very generic class of theories, within a set of assumptions (namely that the particles follow geodesics). For those who believe in the CCC strongly, our results can provide useful fundamental constraints on the type of deviations from Kerr that are possible and still compatible with the CCC. In particular, we show that, generically, the CCC is preserved. For the agnostics, our results give useful information about whether to expect naked singularities in our universe in other, more generic theories of gravity.

\section{Setup: quasi-Kerr}
We assume that GR is modified in the strong field regime, and that compact configurations will suffer parametrically small corrections with respect to their GR counterparts.
Accordingly, the axisymmetric, stationary BH solution is no longer Kerr but an object which deviates slightly from this geometry. We will take it to be of the form~\cite{Johannsen:2011dh,2013ApJ...773...57J,Bambi:2013eb,Cardoso:2014rha,Berti:2015itd},
\begin{eqnarray}
ds^2&&=-\left(1-\frac{2Mr}{\Sigma}\right)(1+\epsilon h_{tt})dt^2+\frac{\Sigma}{\Delta}(1+\epsilon h_{rr})dr^2\nonumber\\
&&+\Sigma(1+\epsilon h_{\theta \theta}) d\theta^2 -\frac{4aMr\sin^2\theta}{\Sigma}(1+\epsilon h_{t\phi})dtd\phi\nonumber\\ 
&&+\left[r^2+a^2+\frac{2a^2Mr\sin^2\theta}{\Sigma}\right]\sin^2\theta(1+\epsilon h_{\phi \phi}) d\phi^2\,, \label{metricPert}
\end{eqnarray}
where $\Sigma\equiv r^2+a^2\cos^2\theta,\,\Delta=r^2-2Mr+a^2$. Here, $\epsilon>0$ parametrizes the deviations from Kerr (for $\epsilon=0$ the above metric reduces to Kerr in standard Boyer-Lindquist coordinates) 
and all the $h_{ij}(r,\theta)$ are arbitrary.

In addition, we assume the following:
\begin{itemize}

\item Free particles follow geodesics in the spacetime \eqref{metricPert}. This is a strong assumption, but since the theory from which the background is derived is unknown, it is a necessary one.
In all calculations we will particularize to the most ``dangerous'' particles, which are null (since their angular momentum /energy ratio is largest); we suspect -- but lack proof -- that for null particles
our results may be common to a wide class of theories without the geodesy assumption.

\item Motion is restricted to the equatorial plane. This assumption is presumably not very restrictive, particles on the equator
carry larger angular momentum and are expected to be the most relevant in the context that we discuss in this work.

\item All the functions $h_{ab}$ are regular everywhere outside the horizon and $\epsilon$ is a small parameter.
This condition ensures that nothing dramatic happens to the metric coefficients, $1+\epsilon h_{ab}\neq 0$. In other words, the functions $h_{ab}$ are smooth and never ``too'' large. 

\item All the metric functions $h_{ab}$ decay faster than $1/r^2$. This ensures that $M, a$ are still the mass and angular momentum of the spacetime with extra mild assumptions on the background theory.

\end{itemize}

In this setup, it is easy to see that there is an event horizon for $g^{rr}=0$~\cite{Johannsen:2013rqa}, and that therefore in these coordinates it sits at $r_+=M+\sqrt{M^2-a^2}$. The extremal $a=M$ limit is of particular interest for us here, since at the perturbative level that we're interested in, it is this point that defines whether or not the spacetime can become super-extremal through the absorption of particles. In this limit, the inverse metric yields 
\be
g^{tt}\sim \frac{4M^2}{4\epsilon(h_{\phi\phi}-2h_{t\phi}+h_{tt})-\delta^2}\,,
\ee
where we are evaluating the metric at $r=M+\delta$ and we neglected terms of order $\epsilon^2$. 
A similar expression holds for $g^{t\phi}$.
For example, for $h_{tt}=h_{\phi\phi}=0$ we find
\be
g^{tt}=-\frac{r(2M^3+M^2r+r^3)}{4\epsilon h_{t\phi}(2+\epsilon h_{t\phi})M^4+M^2r^2-2Mr^3+r^4}\,,
\ee
which diverges at
\beq
r&=&M\frac{1+\sqrt{1\pm8i\sqrt{\epsilon h_{t\phi}(2+\epsilon h_{t\phi})}}}{2}\nonumber\\
&\sim& M+ 2\sqrt{2}M\sqrt{-\epsilon h_{t\phi}}\,,
\eeq
unless $\epsilon h_{t\phi}$ is positive. If the inverse metric diverges at some point outside the horizon, it is easy to verify that certain invariants, such as the Ricci scalar, will also diverge there. Thus, in order to avoid unphysical behavior, we impose that
\be
\Upsilon\equiv -\epsilon\left(h_{\phi\phi}-2h_{t\phi}+h_{tt}\right)>0\,,\label{reg_cond}
\ee
close to the horizon of extremal BHs. We are not including the equality. For $h_{\phi\phi}-2h_{t\phi}+h_{tt}=0$, which includes the Kerr geometry, the terms of higher order
would have to be studied. Obviously, in the case of Kerr the equality is allowed.

\subsection{Particle trajectories}
The original Wald construction consists on throwing point-like particles at a spinning BH with mass $M$ and angular momentum $aM$, and computing the final BH mass $M_f$ and spin $M_fa_f$~\cite{Wald:1997wa}.
Wald proved that after absorption, the spacetime still satisfies $a_f/M_f\leq 1$, with equality attained in the extremal limit. Notice that the particles
in this gedanken experiment are assumed to be point-like and infinitesimally light, so that backreaction effects (conservative and dissipative self-force)
can be neglected~\footnote{For finite-size effects see Refs.\cite{Jacobson:2009kt,Barausse:2010ka,Barausse:2011vx,Colleoni:2015afa}, but we will not consider these here.}.
We can then write
\beq
\frac{a_f}{M_f}&=&\frac{a+m L}{(M+m E)^2}\nonumber\\
&\sim& \frac{a}{M}+m\frac{E}{M}\left(\frac{L}{EM}-2\frac{a}{M}\right)\,,\label{eq:limit}
\eeq
where $m$ is the pointlike particle's rest mass, and $E,\,L$ its intrinsic energy and angular momentum (per unit rest mass), respectively. 
Thus, the extremal limit can be beaten if the second term of Eq.~\eqref{eq:limit} is positive, which for a Kerr BH implies that the absorbed particle carries intrinsic angular momentum 
\be
L/EM>2\,.\label{eq_limit}
\ee
In GR, this is impossible~\cite{Wald:1997wa}.
The analysis is straightforward, as it involves only the computation of geodesics and of the turning points of the motion.
Because the most ``dangerous'' particles are those with small rest mass, we will specialize our results to the massless limit, although the conclusions
are quite general. We will focus entirely on equatorial motion, which is expected to be the most relevant for spin-up. In this case, the conserved energy and angular momentum parameters can be defined through
\beq
&& -E=p_t=\frac{\partial \mathcal{L}}{\partial\dot{t}}=g_{tt}\dot{t}+g_{t\phi}\dot{\phi}\,, \label{EConstant}\\
&& L=p_{\phi}=\frac{\partial \mathcal{L}}{\partial \dot{\phi}}=g_{t\phi}\dot{t}+g_{\phi\phi}\dot{\phi}\,,
\eeq
where $ \mathcal{L} $ is the Lagrangian for the particle under study. A closure relation for geodesics is provided by the normalization condition
\begin{eqnarray} 
\mathcal{L}=g_{\mu\nu}\frac{dx^{\mu}}{ds}\frac{dx^{\nu}}{ds}=\frac{1}{m^2}g^{uv}p_\mu p_\nu =-1\label{lagrangian}\,.
\end{eqnarray}
Note that we will eventually particularize to the null case, by letting $m\to 0$. 
The geodesic motion can now be studied. The relevant calculation consists in studying whether particles with dangerously large angular momentum
can reach and be absorbed at the event horizon. We will first prove a useful bound on the angular momentum of the incoming particle.
With the relations above, we get
\begin{equation} \label{EqEnergy}
E=\frac{Lg^{t\phi}\pm \sqrt{A}}{g^{tt}}
\end{equation}
where 
\begin{equation} 
A = (g^{t\phi}L)^2-g^{tt}\left(m^2+g^{rr}p_r^2+g^{\theta \theta}p_{\theta}^2+g^{\phi \phi}L^2\right)\,.\label{radial}
\end{equation}

Finally, one can also easily show that
\begin{equation} \label{EqEnergy2}
\dot{t} = -\left(\frac{E+g_{t\phi}\dot{\phi}}{g_{tt}}\right)\,.
\end{equation}
The requirement that $\dot{t}>0$ for all future directed timelike motions outside the horizon yields the constraint
\begin{equation} 
\label{Energy3}
E \geq \frac{g^{t\phi}}{g^{tt}}L\,.
\end{equation}
%

\subsection{Preservation of the CCC}
We now show that, generically, but in the context of infalling point particles, the CCC is preserved.
Our working hypothesis is that the full spacetime metric is a small deviation from Kerr, i.e., $\epsilon$ is a small parameter. Thus, the critical angular momentum $L$ at extremality can be expanded as
\be
L=2M+\gamma\,,
\ee
with $\gamma$ a small quantity.
The radial equation of motion \eqref{radial} then yields
\beq
p_r^2&=&\frac{4\gamma r}{(r-1)^3}+\frac{2r}{(r-1)^2}+\frac{r^2}{(r-1)^2}+\frac{\epsilon}{r(r-1)^4}\nonumber\\
&\times&\left[8h_{tt}-8(2+r+r^2)h_{t\phi}+4(2+r+r^3)h_{\phi\phi}+r\left(r(2-3r+r^3)h_{rr}\right.\right.\nonumber\\
&&\left.\left.-(r^4+r^2-6r-4)h_{tt}\right)\right]
\eeq

Expanding the radial equation of motion for very small $\gamma$, one gets the turning point $r_t=M+\delta_t$ at
\be
\delta_t=\frac{2\sqrt{\Upsilon}}{\sqrt{3}}-\frac{4}{9}\epsilon(h_{t\phi}+h_{\phi\phi}-2h_{tt})+{\cal O}(\epsilon^{3/2})\,,
\ee
Because of the regularity condition on the metric, expressed by inequality \eqref{reg_cond}, then $\delta_t$ is always positive 
for any sufficiently small $\gamma$, {\it unless} $\Upsilon=0$, a special case which will be dealt with below. In other words, there is always a turning point outside the horizon and particles with too large angular momentum are not absorbed. A similar result holds for finite $\delta_t > \epsilon$. We have also verified these results numerically for different classes of metric perturbations of monomial type, $h\sim 1/r^b$.
 
For example, for the particular case where $h_{tt}=h_{\phi\phi}=0$, we find
%
\be
\dot{r}^2=\frac{(1+\epsilon h_{rr})r^3\left(r^3+rM^2+2M^3-4LM^2(1+\epsilon h_{t\phi})-(r-2M)L^2\right)}{(r-M)^2\left(4\epsilon M^4 h_{t\phi}(2+\epsilon h_{t\phi})+r^2(r-M)^2\right)}\,.
\ee
%
In this case, one can solve for a turning point at $r_t$ in a simple form, by computing the angular momentum parameter $L$ of a co-rotating geodesic (this can be seen to be co-rotating by checking its value at large distances)
%
\be
L=\frac{2M^2+2M^2\epsilon h_{t\phi}-\sqrt{r^4-2Mr^3+M^2r^2+8\epsilon M^4 h_{t\phi}+4M^4\epsilon^2h_{t\phi}^2}}{2M-r}\,.
\ee
%
Thus, close to the horizon of an extremal BH, $r=M$, one finds 
\beq
L=2M\left(1+\epsilon h_{t \phi}-\sqrt{\epsilon h_{t\phi}(2+\epsilon h_{t\phi})}\right)<2M\,.
\eeq

To sum up, spacetimes which differ from Kerr by parametrically small deviations cannot, generically, be spun up past extremality
by point particles. Horizons are not destroyed and the CCC is preserved~\footnote{This intriguing result does {\it not} contradict a recent claim that {\it regular} BHs can be destroyed through the same gedanken experiment of throwing point particles~\cite{Li:2013sea}; firstly, because Ref.~\cite{Li:2013sea} considers finite-mass particles, while neglecting back-reaction effects. Secondly, even if regular black holes can be destroyed,
they are not harbouring a singularity. In that sense, should the results of Ref.~\cite{Li:2013sea} be correct in the limit that the particle is massless, it would be one more evidence that the central singularity indeed plays an important role in the protection of the surrounding event horizon.}.

\subsection{Exceptions: JP and CPR metrics\label{secexception}}
There are proposals for spacetimes which also differ from Kerr by small quantities, but allowing the perturbations $h_{rr}, h_{rr}, etc$
to diverge at the horizon of the ``background'' Kerr spacetime. These spacetimes were first introduced by  Johanssen and Psaltis (JP)~\cite{Johannsen:2011dh,2013ApJ...773...57J} and later generalized in Ref.~\cite{Cardoso:2014rha} (CPR).
The CPR metric is given by 
\begin{eqnarray}
ds^2&&=-F(1+h^{t})dt^2-2a\sin^2\theta(H-F(1+h^t))dtd\phi\nonumber\\
&&+\frac{\Sigma}{\Delta+a^2\sin^2\theta h^r}(1+h^{r})dr^2+\Sigma d\theta^2\nonumber\\
&+&\sin^2\theta\left(\Sigma+a^2\sin^2\theta(2H-F(1+h^t))\right) d\phi^2\,, \label{metricCPR}
\end{eqnarray}
where $F=1-2Mr/\Sigma$ and $H\equiv \sqrt{(1+h^r)(1+h^t)}$. The functions $h^r, h^t$ are assumed to be radial functions, otherwise free, but typically taken to have an expansion of the form
\begin{align}
h^r = \sum \limits_{k=0}^{+\infty} \left(e_{2k}^r+e_{2k+1}^r\frac{Mr}{\Sigma}\right)\left(\frac{M^2}{\Sigma}\right)^k \label{hr}\\
h^t = \sum \limits_{k=0}^{+\infty} \left(e_{2k}^t+e_{2k+1}^t\frac{Mr}{\Sigma}\right)\left(\frac{M^2}{\Sigma}\right)^k \,,
\end{align}
The CPR metric reduces to Kerr metric in standard Boyer-Lindquist coordinates when we set $h^r=h^t=0$. For the sake of simplicity, we are going to work again in the equatorial plane only.

It is straightforward to show via explicit numerical examples that these spacetimes {\it can} be spun-up past extremality
by point particles following geodesics. This seems to be tied up with the fact that the perturbations are {\it not} parametrically small
for nonzero $h^r$, i.e., they cannot be expressed as regular, small deviations from Kerr.
For example, for $h^t=0, h^r = 0.1M^2/r^2, a=0.956651830214 M$ and $L/M=2+0.001$ we get $L/ME-2a/M=0.0877$. Notice that the value
of rotation was chosen to make it an extremal BH. In view of inequality
\eqref{eq_limit} this is then a spacetime that can be spun-up past extremality.

Perhaps more curious is that we also found that $h^r=0$ leads to a similar result, and this can be proven easily. 
In this case, the inverse metric reads
%
\beq
g^{tt}=\frac{-2(1+h^t)M^3+(1+h^t-2\sqrt{1+h^t})M^2-r^3}{(1+h^t)(M-r)^2r}\,,\nonumber
\eeq
%
and is finite and regular everywhere. Following the procedure and notation of the last section we also find
%
\beq
p_r^2=-\frac{(\delta+1)(\delta^2(-1+h^t)+3\delta(-1+h^t)+2h^t)}{\delta^3}\,.
\eeq
Therefore, one can conclude that the turning points are given by
\beq
\delta=\frac{3-3h^t\pm\sqrt{9-10h^t+{h^t}^2}}{2(-1+h^t)}, 
\eeq
hence $\delta$ can be made negative if $h^t<0$, a result which we verified numerically. 

The $h^r=0$ case happens to belong to the category of spacetimes described earlier, with
\beq
\epsilon h_{tt}&=&h^t\,,\\
\epsilon h_{t\phi}&=&\frac{2Mh^t-r(h^t+1)+r\sqrt{1+h^t}}{2M}\,,\\
\epsilon h_{\phi\phi}&=&a^2\frac{h^t(2M-r)+2r\left(\sqrt{h^t+1}-1\right)}{r^3+a^2(2M+r)}\,.\\
\eeq
However, there is no contradiction with our previous claims, as it so happens (coincidentally) that for these perturbations
$\Upsilon=0$ at the horizon to order $\epsilon$. 

\section{Conclusions}
Our initial goal was to show that very large classes of modified geometries, parametrically close to the Kerr geometry,
would lead to violations of cosmic censorship. Instead, it turns out that cosmic censorship is still preserved, for most
of the geometries considered, with few very special exceptions.
Together with previous results on Kerr~\cite{1974AnPhy..82..548W} or charged BHs~\cite{Gao:2012ca,Hod:2013vj}, higher-dimensional BHs~\cite{BouhmadiLopez:2010vc} and BHs in non-asymptotically flat spacetimes~\cite{Rocha:2011wp,Rocha:2014jma,Zhang:2013tba,Gwak:2015fsa},
our results indicate that destroying event horizons is a difficult task. We remain as ignorant as before on the precise reasons for this.

\section{Acknowledgements}
%
We thank Jorge Rocha and Masashi Kimura for useful comments and feedback.
V.C. acknowledges financial support provided under the European
Union's FP7 ERC Starting Grant ``The dynamics of black holes: testing
the limits of Einstein's theory'' grant agreement no. DyBHo--256667
and H2020 ERC Consolidator Grant 
``Matter and strong-field gravity: New frontiers in Einstein's theory'' grant agreement no. MaGRaTh--646597.
This research was supported in part by the Perimeter Institute for
Theoretical Physics. Research at Perimeter Institute is supported by
the Government of Canada through Industry Canada and by the Province
of Ontario through the Ministry of Economic Development $\&$
Innovation.
This work was supported by the H2020-MSCA-RISE-2015 Grant No.
StronGrHEP-690904.
%

\end{document}